\documentclass[preprint,floatfix,prl,showpacs,amsmath,amssymb,superscriptaddress]{revtex4}

\usepackage{graphicx}
\usepackage{dcolumn}
\usepackage{amssymb}
\usepackage{amsmath}
\usepackage{color}
\usepackage{bm}
\usepackage{verbatim}
\newcommand{\ua}{\uparrow}
\newcommand{\da}{\downarrow}

\def\braket#1{\mathinner{\langle{#1}\rangle}}
\def\bra#1{\left\langle{#1}\right|}
\def\ket#1{\left|{#1}\right\rangle}

\begin{document}
 
\title{Suppression of Zeeman gradients by nuclear polarization in double quantum dots}

\author{S. M.~Frolov}
\affiliation{Kavli Institute of Nanoscience, Delft University of Technology, 2600 GA Delft, The Netherlands}
\author{J.~Danon}
\affiliation{Dahlem Center for Complex Quantum Systems, Freie Universit\"{a}t Berlin, 14195 Berlin, Germany}
\author{S.~Nadj-Perge}
\affiliation{Kavli Institute of Nanoscience, Delft University of Technology, 2600 GA Delft, The Netherlands}
\author{K.~Zuo}
\affiliation{Kavli Institute of Nanoscience, Delft University of Technology, 2600 GA Delft, The Netherlands}
\author{J. W. W.~van Tilburg}
\affiliation{Kavli Institute of Nanoscience, Delft University of Technology, 2600 GA Delft, The Netherlands}
\author{V.~S.~Pribiag}
\affiliation{Kavli Institute of Nanoscience, Delft University of Technology, 2600 GA Delft, The Netherlands}
\author{J. W. G.~van den Berg}
\affiliation{Kavli Institute of Nanoscience, Delft University of Technology, 2600 GA Delft, The Netherlands}
\author{E. P. A. M.~Bakkers}
\affiliation{Kavli Institute of Nanoscience, Delft University of Technology, 2600 GA Delft, The Netherlands}
\affiliation{Department of Applied Physics, Eindhoven University of Technology, 5600 MB Eindhoven, The Netherlands}
\author{L. P.~Kouwenhoven}
\affiliation{Kavli Institute of Nanoscience, Delft University of Technology, 2600 GA Delft, The Netherlands}

\begin{abstract}
We use electric dipole spin resonance to measure dynamic nuclear polarization in InAs nanowire quantum dots. The resonance shifts in frequency when the system transitions between metastable high and low current states, indicating the presence of nuclear polarization. We propose that the low and the high current states correspond to different total Zeeman energy gradients between the two quantum dots. In the low current state, dynamic nuclear polarization efficiently compensates the Zeeman gradient due to the $g$-factor mismatch, resulting in a suppressed total Zeeman gradient. We present a theoretical model of electron-nuclear feedback that demonstrates a fixed point in nuclear polarization for nearly equal Zeeman splittings in the two dots and predicts a narrowed hyperfine gradient distribution.
\end{abstract}


\maketitle

Hyperfine interaction couples electron spin to nuclear spins enclosed by the electron's wave function. In the context of spin qubits in III-V semiconductors, the most prominent effect of this interaction is that fluctuating nuclear spins cause electron spin dephasing \cite{pettascience05, Koppens2008, Nadj-Perge2010a}. Interestingly, ideas for suppressing nuclear spin fluctuations also rely on the same hyperfine interaction, since electron spin transport can lead to dynamical nuclear polarization (DNP) in quantum dots \cite{onoprl04}. An experimental manifestation of DNP is a hysteretic current in the spin blockade regime  \cite{Koppens2005,  RudnerPRL07, churchillnatphys09, kobayashiprl11}. Sometimes, including in the present work, the hysteresis could be extended to high magnetic fields, suggestive of a large degree of nuclear polarization \cite{baughprl07, pfundprl07}.

In this report, we study hysteretic spin blockade in InAs nanowire quantum dots using electric dipole spin resonance (EDSR) spectroscopy \cite{Vink2009, Schreiber2011, Nadj-Perge2012}. Surprisingly, the degree of polarization deduced from EDSR does not exceed a few milliTesla, much smaller than the hysteresis range. We explain this apparent contradiction by nuclear fields compensating the natural Zeeman energy difference between the two quantum dots caused by the mismatch of their $g$-factors \cite{SchroerPRL11}. In this case one of the double dot states ($T_0$) is blocked leading to a reduced current. We support this idea by analytical and numerical calculations of spin blockade transport in the presence of hyperfine and spin-orbit interactions.

From our model we deduce a narrowing of the hyperfine gradient distribution to a few percent of the unpumped distribution width. This finding is especially relevant for two-electron singlet-triplet qubits, where the hyperfine {\it gradient} is the source of dephasing \cite{taylor:natphys05}. We predict an order of magnitude enhancement in the coherence time due to gradient suppression induced by spin blockade transport. This is an alternative route to $T_2^*$ enhancement compared with nuclear spin pumping by pulsing the double dot through an $S$-$T$ transition \cite{FolettiNatPhys09, BluhmPRL10, BluhmNatPhys11}.

We present data from two devices that were studied in two previous publications \cite{Nadj-Perge2010a, nadj-perge2010}. InAs nanowires, 40--80 nm in diameter, are deposited on top of five narrow bottom gates which are then used to define few electron double quantum dots. The nanowires are contacted by Ti/Al leads to measure electron transport through the system. All measurements are performed at $T=250$--$300$~mK in ${}^3$He refrigerators.

\begin{figure}[t]
\includegraphics[width=8.5cm]{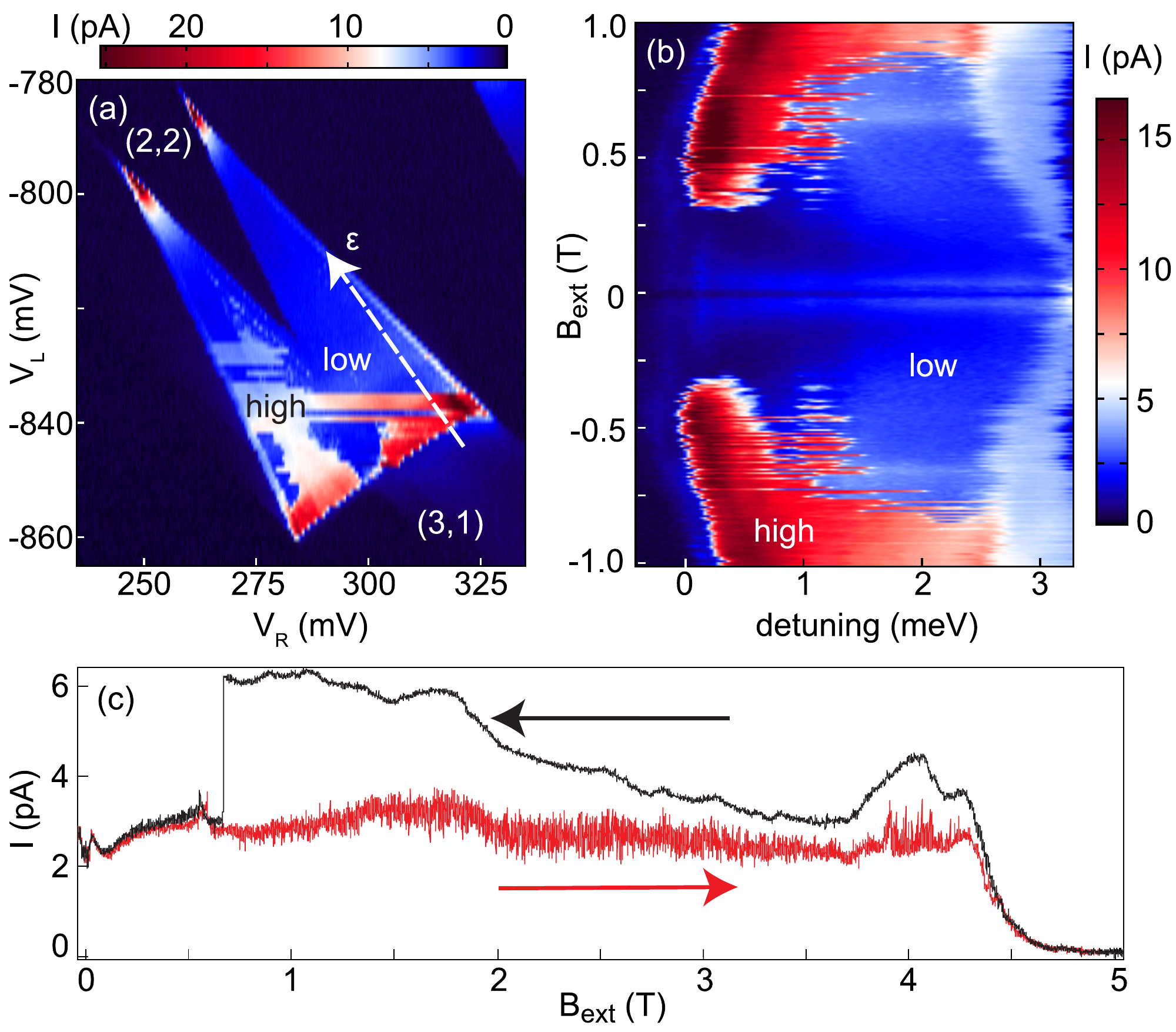}
\caption{(a) Charge stability diagram near the (3,1)$\rightarrow$(2,2) transition ($V_\text{dc}=7$ mV, $B_\text{ext} = 0.79$ T). Low and high current states are labeled, the dashed arrow indicates the detuning axis $\varepsilon$. (b) Dependence of the leakage current on magnetic field and detuning. The detuning is swept along the line in panel (a), and the field is stepped after each sweep. For (a) and (b) the sweep direction is from left to right. (c) Magnetic field retrace at fixed detuning $\varepsilon \approx 3.5$ meV, the sweep directions are indicated with arrows. The current becomes zero above 4 T since the double dot then shifts into a Coulomb blocked state due to the Zeeman shift of $T_+$(1,1). The same device was previously studied in \cite{nadj-perge2010}.}
\end{figure}

We tune the double dots to the so-called strong coupled spin blockade regime, in which the current exhibits a dip rather than a peak at zero magnetic field due to an interplay between tunnel coupling, hyperfine and spin-orbit interactions \cite{nadj-perge2010, pfundprl07}. Fig.\ 1(a) shows the double dot current in the vicinity of a spin-blocked charge degeneracy point, which appears as a double-triangle shape when the left and right gates are swept at a finite d.c.\ voltage bias $V_\text{dc}$ across the double dot. Inside the triangles, one can see sudden transitions between a low and a high current state. The difference in current between the two states exceeds, for some settings, an order of magnitude (1--2 pA vs.\ 10--20 pA). This switching is only visible {\it within} the boundaries of the triangles. The boundaries themselves remain fixed in gate voltage. This indicates that the jumps in current do not originate from charge switches in the vicinity of the double dot.

The appearance of the two current states is strongly influenced by an applied magnetic field $B_\text{ext}$ [Fig.\ 1(b)]. At zero field we always observe a stable low current because we are in a zero-field dip characteristic of all strongly coupled quantum dots \cite{Koppens2005, nadj-perge2010, danonprb09}. In the absence of hysteresis this dip is hundreds of mT wide with current increasing smoothly with field. However, in the regime where we observed the switching, the current drops when the field increases beyond 10--20 mT, and the system enters a metastable low current state \cite{pfundprl07}. At higher fields, both low and high current states are observed, as well as sharp transitions between the two.

In Fig.\ 1(c) we present an example of a hysteretic current trace. We see that the low current state can be ``dragged'' up to very high fields (over 4 T). The reversed sweep shows a distinct high current state down to $\sim 0.7$ T, where the current switches. If the magnetic field is fixed to a value inside the hysteretic regime, the double dot may suddenly switch after minutes or seconds, or remain in either of the states for as long as hours (see supplementary material for time-dependent measurements). We note that the higher noise observed in the low current state does not represent typical behavior, often the current fluctuations are larger in the high current state.

Electron spin-nuclear spin feedback is known to exhibit complex dynamics, including hysteresis \cite{Vink2009,RudnerPRL07}, multi-stabilities \cite{Danon2009}, and fast switching between different stable states \cite{Koppens2005, RudnerPRB11}. This suggests that the observed switching and hysteresis might be due to DNP. Earlier experiments attempted to extract the degree of nuclear polarization directly from the hysteresis curves \cite{pfundprl07, baughprl07}. In this interpretation, the maximum nuclear polarization is simply given by the size of the hysteresis loop. In our case, Fig.\ 1(c) would present a contradiction to such an interpretation. The field range of hysteresis exceeds 3 T, which is a few times larger than the effective field corresponding to full polarization for InAs ($\approx$~1.0--1.5 T, depending on the effective electronic $g$-factor). This shows that a straightforward analysis of the hysteresis curves does not provide an estimate for the maximum nuclear fields in the dots.

\begin{figure}[t]
\centering
\includegraphics[width=8.5cm]{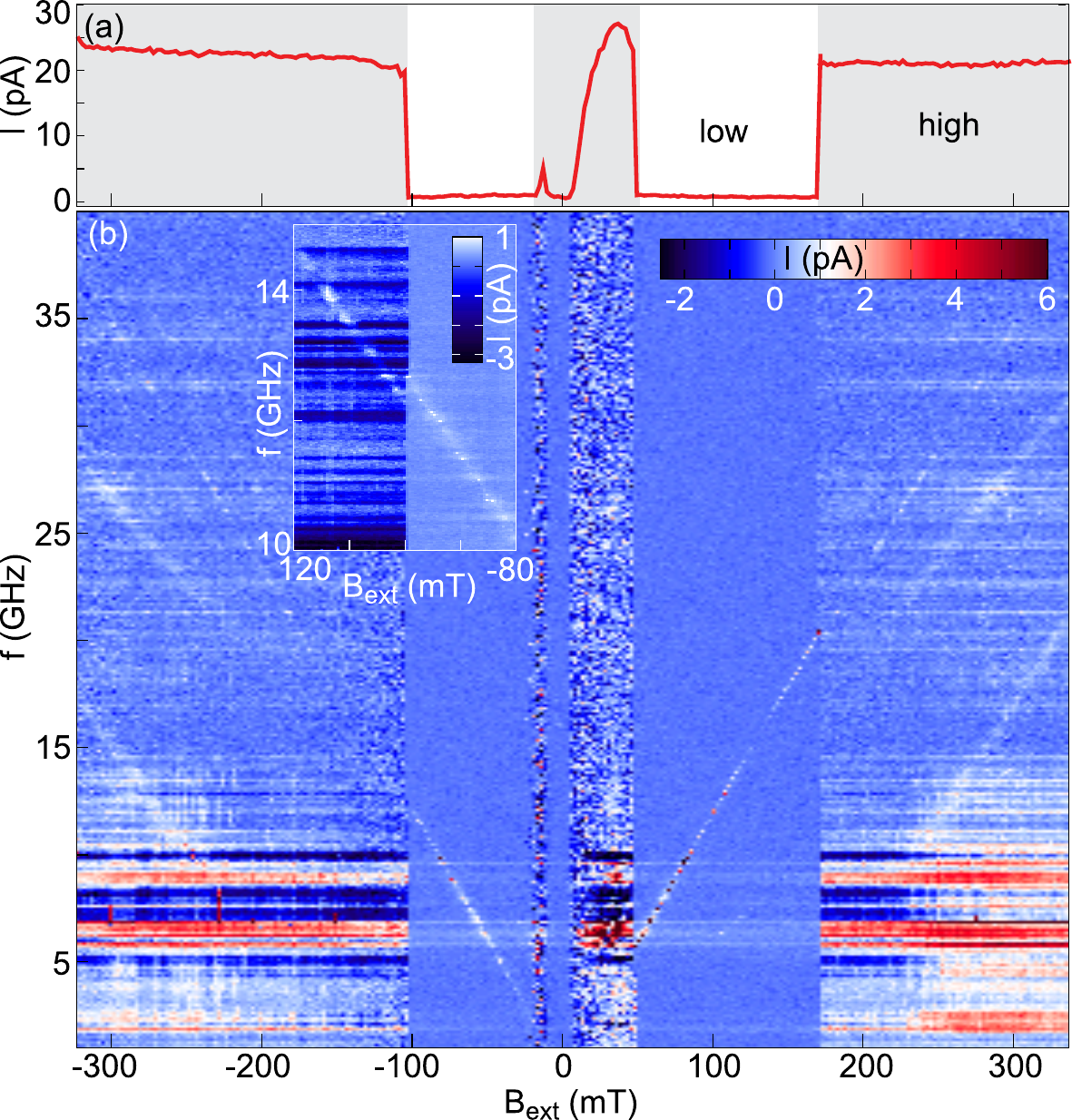}
\caption{(a) Double dot current as a function of magnetic field (device 2, $V_\text{dc}=7.5$ mV, and $\varepsilon \approx 2.5$ meV). The gray and white background indicates the high and low current states respectively. (b) The frequency of the a.c.\ voltage on a local gate is swept, while the magnetic field is stepped. At each point, we measure 50 ms with a.c.\ excitation on the gate and then 1 s without. The difference in the two measured currents is plotted. This procedure also avoids dragging of the spin resonance. The inset shows a zoom-in on the spectral line at the transition from the high to low current state. Low a.c.\ power is used to minimize switches between metastable states induced by the driving. The horizontal lines at fixed frequency correspond to photon assisted tunneling enhanced by cavity modes in the fridge. This is the same device as studied in \cite{Nadj-Perge2010a}.}
\end{figure}

We are able to determine the nuclear polarization directly, by probing the effective Zeeman splitting of the electrons. To this end, a continuous wave GHz-frequency electric field is applied to one of the gates. The oscillating electric field drives transitions between the Zeeman-split spin-orbital eigenstates of electrons. When the a.c.\ frequency matches the Larmor precession frequency $f$ in one of the dots, additional transitions within the (1,1) manifold are induced and extra current flows through the double dot. In the presence of DNP, the Larmor frequency is given by $2\pi f = |g\mu_B (\vec B_\text{ext} + \vec B_N) |$, where $g$ is the effective $g$-factor in the quantum dot, $\mu_B$ is the Bohr magneton, $\vec B_N$ is the effective nuclear field, and we have set $\hbar = 1$ for convenience. Therefore, a finite $\vec B_N$ will reflect in a shift of $f$. 

The EDSR spectroscopy is performed on a second device, the spectrum is shown in Fig.\ 2(b) and the high and low current states in Fig.\ 2(a). In the low current state we fit the observed resonance frequency to $2\pi f = |g\mu_B B_\text{ext}|$, yielding $g = 8.7 \pm 0.1$. Similar $g$-factors were measured in this device in a regime where hysteresis was not observed \cite{Nadj-Perge2010a}. The scenario proposed in \cite{pfundprl07}, in which DNP compensates the external field fully in the low current state to a nearly zero total Zeeman energy, can thus again be ruled out. We also note that the extent of the low current state is asymmetric with respect to the zero field axis \cite{RudnerRashbaPRB11}. This effect is further explored in the supplementary material.

A second important observation is that only a single EDSR resonance is observed in the low current state, while in the high current state we see multiple EDSR resonances. In strongly coupled quantum dots we {\it do} expect multiple resonances corresponding to various transitions within the manifold of (0,2) and (1,1) singlet and triplet states \cite{Nadj-Perge2012}. The observation of only a single resonance in the low current state is thus surprising. We at least expect to see two EDSR lines due to a $g$-factor difference between the two dots \cite{Nadj-Perge2010a}.

We also detect a 0.3 GHz shift in the primary resonance frequency at the transition boundary [Fig.\ 2(b), inset]. This suggests that the two current states {\it do} have different nuclear polarizations, the difference being however rather small (at least in one of the dots). How could a small change in polarization alter the current through the double dot by an order of magnitude? And why is only a single EDSR resonance observed in the low current state? In what follows we propose an explanation based on the {\it gradient} in the $z$-projection of the nuclear fields $\Delta B_N^z$ over the two dots \cite{kobayashiprl11}. We stress that the physics presented here is especially relevant for materials with the strong spin-orbit interaction which couples polarized triplets to singlets.

Let us first consider the basics of transport through a spin-blocked (1,1)$\rightarrow$(0,2) transition in the presence of spin-orbit interaction and in the strong interdot coupling regime. An electron enters the left quantum dot and forms one of the four (1,1) states with an electron on the right dot. Out of the four states, only the (1,1) spin singlet is tunnel coupled to the (0,2) singlet due to spin selection rules [the (0,2) triplet states are at too high energy to play a role]. In the absence of a magnetic field, this results in three blocked states and therefore very low current, proportional to the residual escape rate out of the blocked states. If a finite magnetic field is applied, the spin-orbital $T_+(1,1)$ and $T_-(1,1)$ are split off by the Zeeman energy $\pm E_Z = \pm \bar g\mu_B B_\text{ext}$ ($\bar g$ being the average $g$-factor of the two dots), see Fig.\ 3(a). Spin-orbit interaction then effectively allows for spin-nonconserving tunneling and couples $T_\pm(1,1)$ to $S(0,2)$, characterized by the energy $t_\text{so}$ \cite{danonprb09}. The remaining blocked state $T_0(1,1)$ forms the bottleneck for transport and the escape rate out of this state determines the current.

A difference $\Delta E_Z$ in effective Zeeman splittings in the two dots mixes $T_0(1,1)$ with $S(1,1)$, thereby unblocking the system at finite magnetic field \cite{Koppens2005}, as indicated in Fig.\ 3(a). $\Delta E_Z$ is contributed to by a difference in the effective $g$-factors of the two dots $\Delta g$, as well as by a nuclear field gradient along the $z$-axis. Since $T_0(1,1)$ constitutes the bottleneck in the transport cycle, a change in $\Delta B_N^z$ (and thus in $\Delta E_Z$) could indeed have a significant effect on the current. We support this statement by performing transport simulations including a Zeeman gradient. Fig.\ 3(b) shows the calculated double dot current versus $B_\text{ext}$, for two values of $\Delta E_Z$. We used a rate equation model that includes the effects of hyperfine and spin-orbit interactions \cite{nadj-perge2010}. A gradient of only a few mT is sufficient to increase the current by almost an order of magnitude. Note that we have previously reported higher-than-expected current levels in the strong coupling regime \cite{nadj-perge2010}. We now propose that this higher current was due to the Zeeman gradient over the two dots which was not included in the model at the time.
\begin{figure}[t]
\centering
\includegraphics[width=8.5cm]{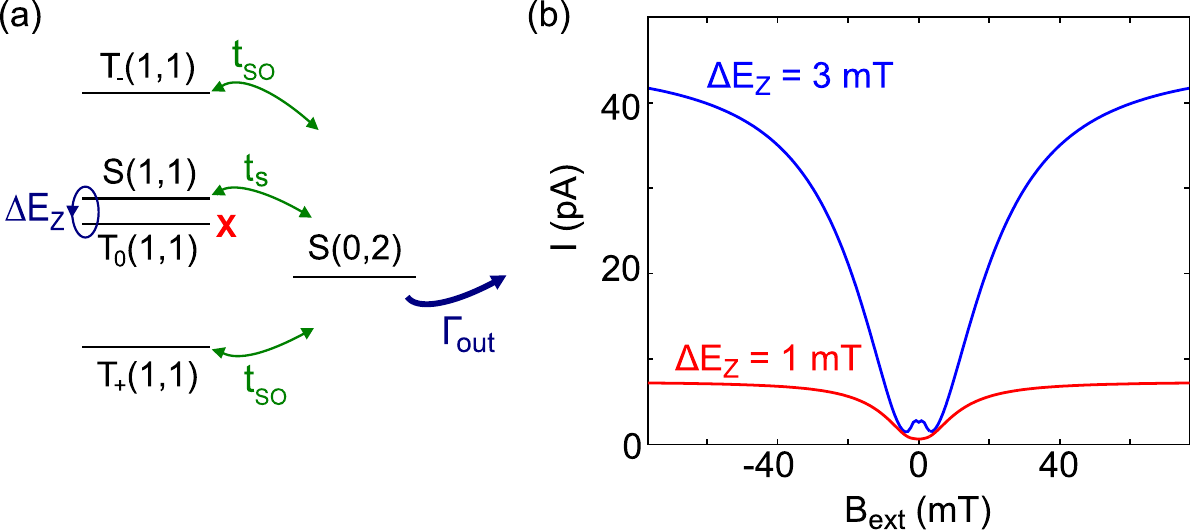}
\caption{(a) Sketch of the spectrum of the five electronic states. The coupling between the different states, as well as the decay rate $\Gamma_\text{out}$ of $S(0,2)$ are indicated. (b) Simulated current in the presence of an external Zeeman gradient of 1 mT (red trace) and 3 mT (blue trace). We used $\varepsilon = \Gamma_\text{out} = 1.5$ meV, $t_s = 100~\mu$eV, $t_\text{so} = 25~\mu$eV, and $\bar g=9$. We averaged over $10^5$ random nuclear fields taken from a normal distribution with $\sqrt{\langle (B_N^{x,y,z})^2 \rangle} = B_N^\text{r.m.s.} = 1$ mT for both dots.}
\end{figure}

In InAs double dots, it has been observed that typically $\Delta g / \bar g =$~1--10 \% \cite{Nadj-Perge2010a, SchroerPRL11}. At applied fields of 100 mT this mismatch would induce a Zeeman gradient of several mT and thus lead to a considerable increase of current due to lifting of the spin blockade [Fig.\ 3(b)]. Low current could arise when a small nuclear field gradient exactly compensates the Zeeman gradient due to $\Delta g$. The low current state thus has $\Delta E_Z \approx 0$. This idea also explains that only a single EDSR resonance is observed in the low-current state [Fig.\ 2(b)] in contrast to Ref.\ \cite{Nadj-Perge2010a} where in the same device two resonances corresponding to two dots were resolved.

\begin{figure}[b]
\centering
\includegraphics[width=8.5cm]{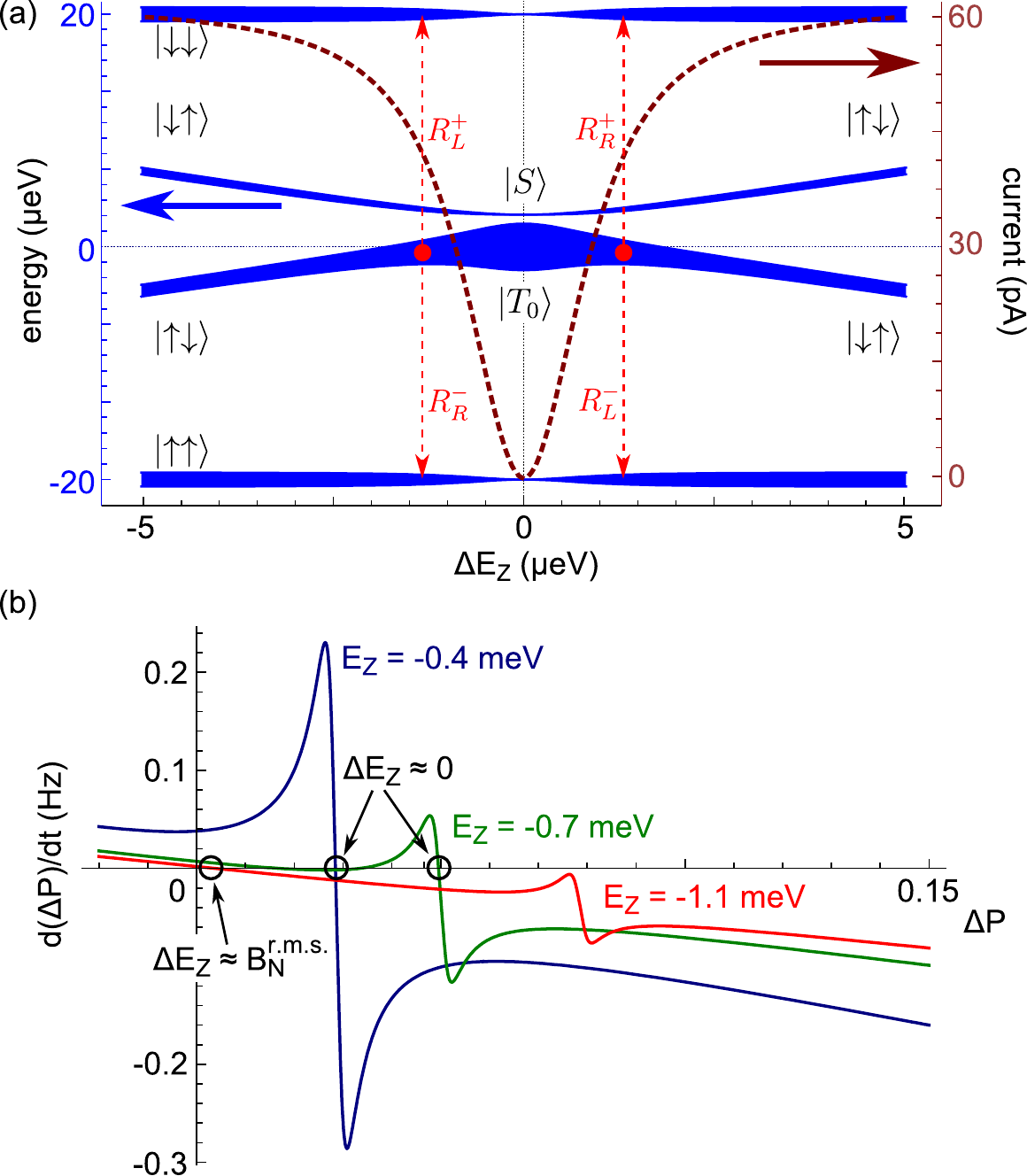}
\caption{(a) Thick blue lines, left scale: Spectrum of the (1,1) states as a function of $\Delta E_Z$ for $E_Z = -20~\mu$eV, $t_s = 60~\mu$eV, $t_\text{so} = 40~\mu$eV, and $\varepsilon = \Gamma_\text{out} = 1.5$ meV. The thickness of the lines indicates the occupation probability of the corresponding eigenstate (4 $\mu$eV = 1). Brown dashed line, right scale: Current through the double dot as a function of $\Delta E_Z$ for the same parameters. Red dashed arrows: Preferred directions of nuclear spin flips close to $\Delta E_Z=0$. (b) $d(\Delta P)/dt$ as a function of $\Delta P$ for three different magnetic fields, negative $E_Z$ corresponds to a positive field. We used the same parameters as at (a), with $AI = 0.7$ meV, $I = \tfrac{1}{2}$, $N = 10^5$, $1/\tau = 3 \cdot 10^{-10}~\mu$eV, and $\Delta g/\bar g = 0.05$. Stable (un)polarized states of the nuclear spin ensembles are indicated.}
\end{figure}

In what follows we show theoretically that electron-nuclear spin flip-flops indeed can drive the double dot towards $\Delta E_Z = 0$ and keep it there. In Fig.\ 4(a) we plot the spectrum of the (1,1) states as a function of $\Delta E_Z$ close to $\Delta E_Z = 0$. The thickness of the lines in the spectrum corresponds to the occupation probabilities of the four states one finds when taking into account the coupling to the decaying $(0,2)$ singlet: at $\Delta E_Z = 0$ the system has one blocked state $\ket{T_0}$ in which it spends all its time. When $\Delta E_Z$ deviates from zero, $\ket{T_0}$ and $\ket{S}$ acquire a spin-orbital $\ua\da$- and $\da\ua$-character and the blockade is lifted. Correspondingly, the double dot current increases as $\Delta E_Z$ moves away from zero.

At $\Delta E_Z = 0$ positive and negative pumping rates are balanced. Hyperfine induced electron-nuclear spin flip-flops can cause transitions from the only occupied state $\ket{T_0}$ to $\ket{\ua\ua}$ and to $\ket{\da\da}$, with equal probabilities. Since $\ket{T_0}$ is an equal superposition of $\ket{\ua\da}$ and $\ket{\da\ua}$, these processes will not lead to a net pumping of the nuclear fields. 

As soon as $\Delta E_Z$ deviates from zero, DNP tries to return the system to $\Delta E_Z = 0$. For example, when $\Delta E_Z$ is positive, the most strongly occupied state $\ket{T_0}$ acquires a $\da\ua$-character and transitions starting from $\ket{T_0}$ have a preferred direction for spin-flips \cite{PhysRevB.81.041304}. This yields DNP with net rates $R_L^-$ and $R_R^+$ [red dashed arrows in Fig.\ 4(a)]: the left dot is negatively pumped, the right dot positively. Both these rates indeed drive the system back to $\Delta E_Z = 0$. Transitions from $\ket{S}$ have the same asymmetry but with opposite sign and counteract this pumping. However, since the occupation probability of $\ket{S}$ is much smaller than that of $\ket{T_0}$, transitions from $\ket{T_0}$ dominate.

This intuitive picture is confirmed by an explicit calculation of all allowed hyperfine flip-flop rates. The derivation involves a few straightforward steps, following the approach of previous works \cite{RudnerPRL07,Danon2009} (see the supplementary material for details). Assuming nuclear spin 1/2 for simplicity, we calculate each separate transition rate using Fermi's golden rule. Summing over all transition rates, we arrive at an equation of motion for the polarization gradient over the dots:
\begin{equation}
\frac{d\Delta P}{dt} = -
\frac{A^2}{4 N^2E_Z^2}\frac{\Gamma_\text{out}t_s^2}{\Gamma_\text{out}^2 + 4\varepsilon^2}
f(\theta)
[\sin 2\theta+2\Delta P]-\frac{\Delta P}{\tau},
\label{eq:pumpdp}
\end{equation}
where $\Delta P = \tfrac{1}{2} (P_L-P_R)$, with the nuclear polarization in left and right dots $-1<P_{L(R)} <1$. $N$ is the number of nuclei in each dot (see supplementary material for a discussion of asymmetric dot sizes). $A$ is the average hyperfine coupling energy ($AI \sim 0.7$ meV for InAs, with $I$ the average total nuclear spin). The angle $\theta$ is defined by $\tan \theta = (\Delta E_Z)(\Gamma_\text{out}^2+4\varepsilon^2)/2 \varepsilon (t_s^2+t_\text{so}^2)$, where $\Delta E_Z = (\Delta g)\mu_B B_\text{ext} + AI(\Delta P)$, and the dimensionless function $f(\theta) \sim 1$ is given in the supplementary material. In deriving (\ref{eq:pumpdp}), we assumed that $|E_Z| \gg \Gamma_a$ for all (1,1) states. We also added a phenomenological nuclear spin relaxation rate $1/\tau \sim 0.1$-$1$~Hz. The average polarization $P = \tfrac{1}{2} (P_L+P_R)$ is not pumped: DNP merely enhances the relaxation rate of $P$.

In Fig.\ 4(b) we plot the pumping curve (\ref{eq:pumpdp}) for $\Delta P$ with realistic parameters, for different magnetic fields. We see that at low fields the system has one single stable state close to the low current point $\Delta E_Z = 0$ (blue curve, $E_Z = -0.4$~meV). At intermediate fields, the high current unpolarized state with $\Delta P \approx 0$ is stable as well (green curve, $E_Z = -0.7$~meV). This bistability can manifest itself in switching and hysteresis. At very high fields, DNP becomes too weak to counteract nuclear spin relaxation and only the high current unpolarized state is stable (red curve, $E_Z = -1.1$~meV). Using parameters from \cite{nadj-perge2010} we estimate the maximum field for which DNP can stabilize the $\Delta E_Z \approx 0$ to be 6 T, which is indeed consistent with Fig.\ 1(c) (see supplementary information for details).

Often, states of nuclear polarization stabilized by DNP exhibit significantly reduced fluctuations. Using the spin-flip rates found from the Fermi's golden rule we evaluate the variance of the gradient distribution in the polarized state as compared to the unpolarized state. We find a relative suppression of the mean-square fluctuations of about $\approx 4\cdot 10^{-3}$ for the parameters of the blue curve in Fig.\ 4(d), and of $\approx 9\cdot 10^{-3}$ for those of the green curve (see supplementary material for details). In the context of two-electron singlet-triplet qubits, this narrowing could lead to an enhancement of the dephasing time $T_2^*$ by more than an order of magnitude.

We thank  M.\ Rudner, H.\ Bluhm, E.\ Rashba, Yu. Nazarov, L.\ Levitov, L.\ Vandersypen, and K.\ Nowack. This work has been supported by NWO/FOM (Netherlands Organization for Scientific Research), by the Alexander von Humboldt Foundation, and through the DARPA program QUEST.

\section{SUPPLEMENTARY INFORMATION}

\begin{figure}[hp]
\includegraphics {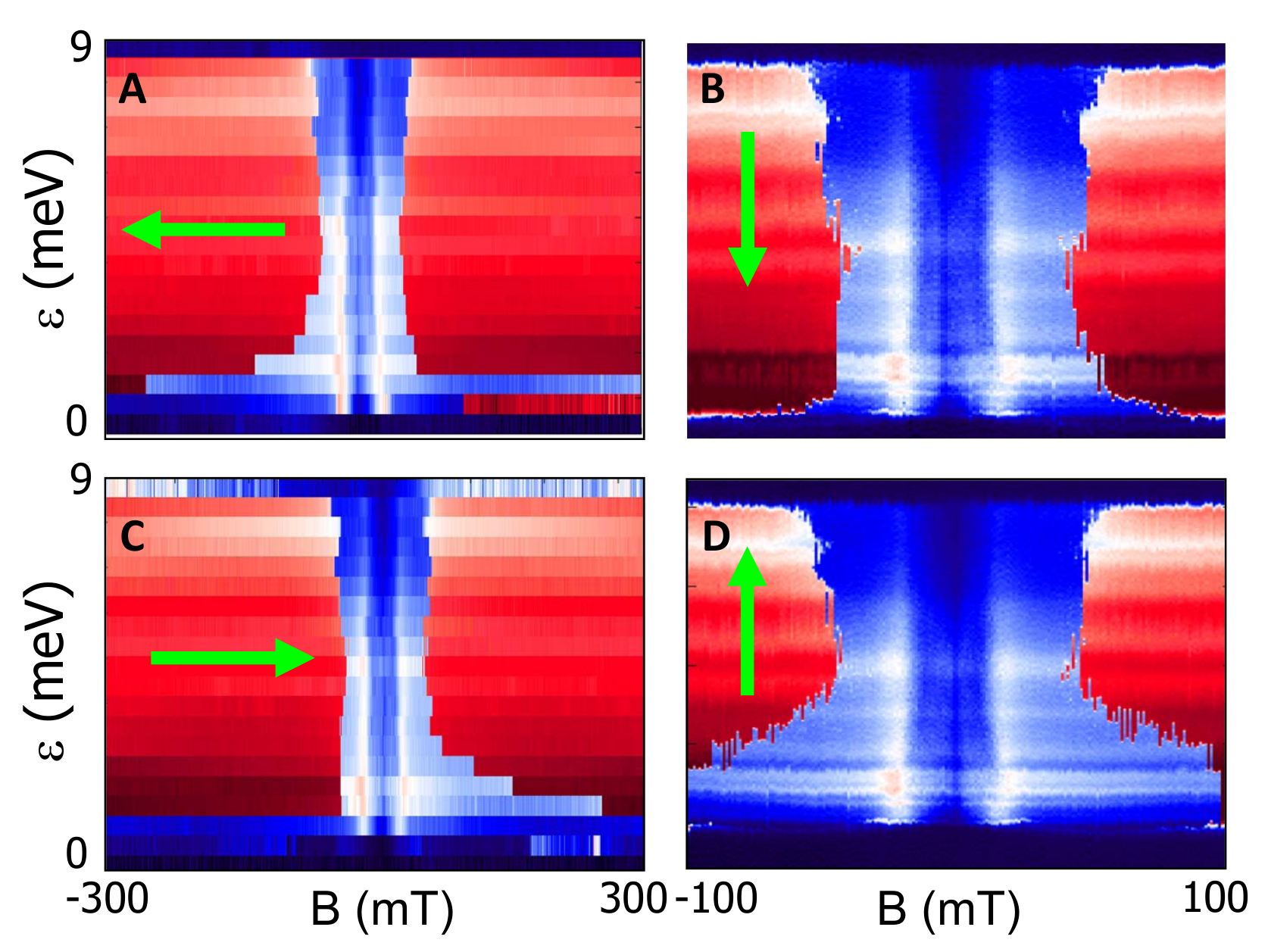}
\caption{Dragging of the low current state depends on the sweep direction. This Figure shows the current in the space of detuning and magnetic field. The detuning cut is through a (1,3)$\to$(2,2) transition in device 1 at $V_\text{dc} = - 9$ mV. Green arrows indicate the sweep direction, the variable along the orthogonal axis is stepped after each sweep. The low current state (blue) occupies a different area in both detuning and field when (a) the field is swept from positive to negative, (b) the detuning is swept from high to low, (c) the field is swept from negative to positive, (d) the detuning is swept from low to high. Panels (a) and (c) are part of the same scan in which the field is swept back and forth. In these panels the system does not leave the low current state for small detuning because the field is not swept high enough to trigger a switch.}
\end{figure}

\begin{figure}[hp]
\includegraphics {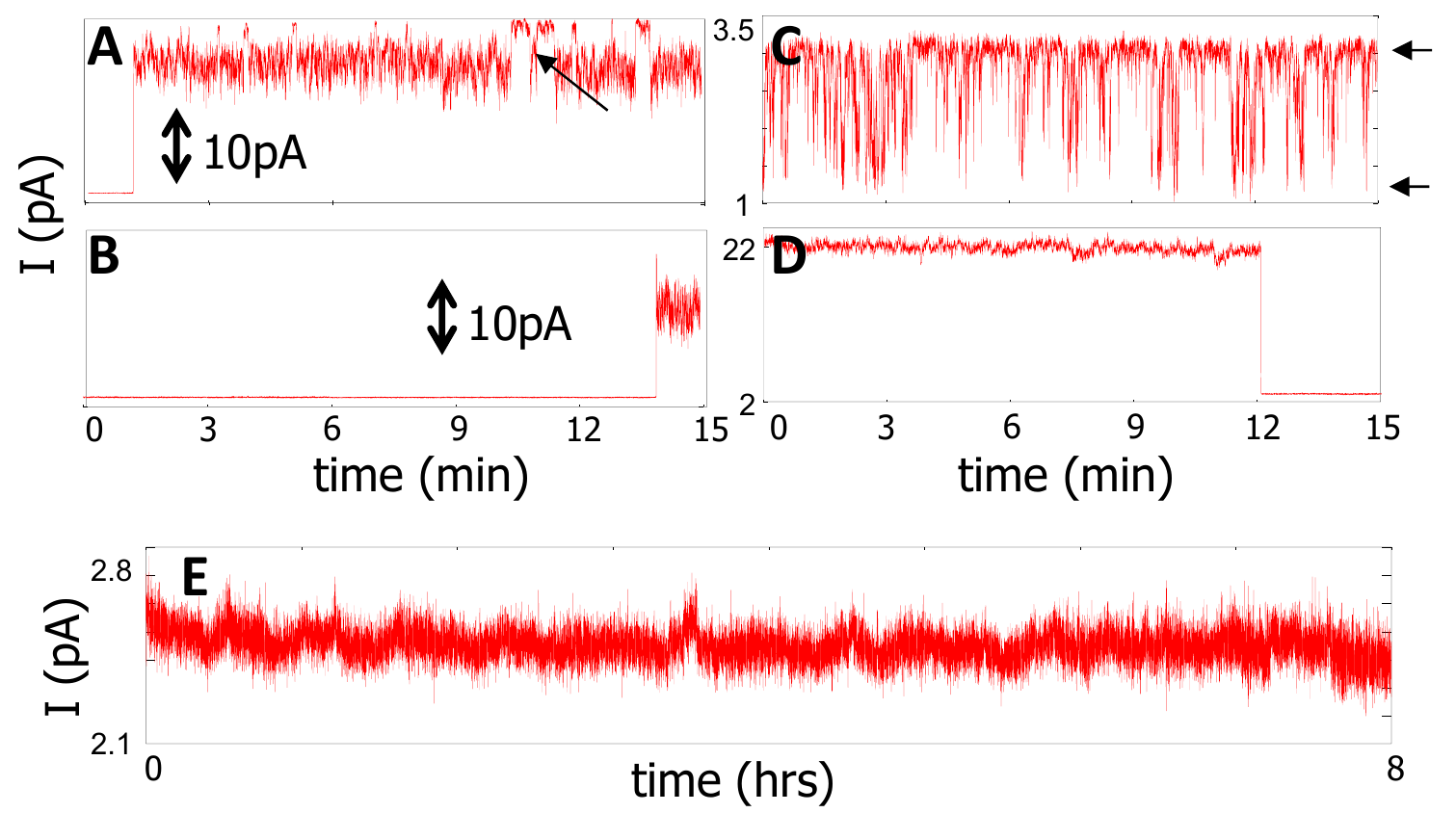}
\caption{The ``waiting game''. Rules: First, the detuning and dc bias are set. Then the field is swept from zero to $B_0$. As soon as $B_0$ is reached the data are recorded as a function of time. (a) and (b): $B_0 = 300$~mT, near the field where the system would switch to the high current state in a typical field sweep [device 1, (1,3)$\to$(2,2)  transition]. Here the system typically switches to the high current state after a time that varies between 1 to 15 minutes. We never observe a gradual current increase between the low and the high current states, the change is always abrupt. If we rapidly turn the dc bias off and then back on, the current would return in the high current state at these settings. In these plots the high current state is characterized by higher fluctuations. Some of the switches are due to charge noise [arrow in panel (a)]. The higher frequency current noise may originate from multiple metastable current states. (c) and (d): Examples of a different behavior. In panel (c) the system switches back and forth between high and low current states (arrows on the right side). These data are taken at the (3,1)$\to$(2,2) transition in the vicinity of a high current state ``island'', i.e., a region of high current state surrounded by a low current state, similar to the one shown in Fig.\ S4(b), at a detuning of $\approx 5$~meV and $B_0 \approx 0.5$~T. In panel (d), the field is swept from high to low to reach $B_{0} = 80$~mT, such that the initial state is the high current state. Panel (e) is an example of low current state persistent for hours. These data are taken in the center of the field hysteresis loop [device 1, (1,3)$\to$(2,2) transition, $B_0 = 200$~mT].}
\end{figure}

\begin{figure}[hp]
\includegraphics {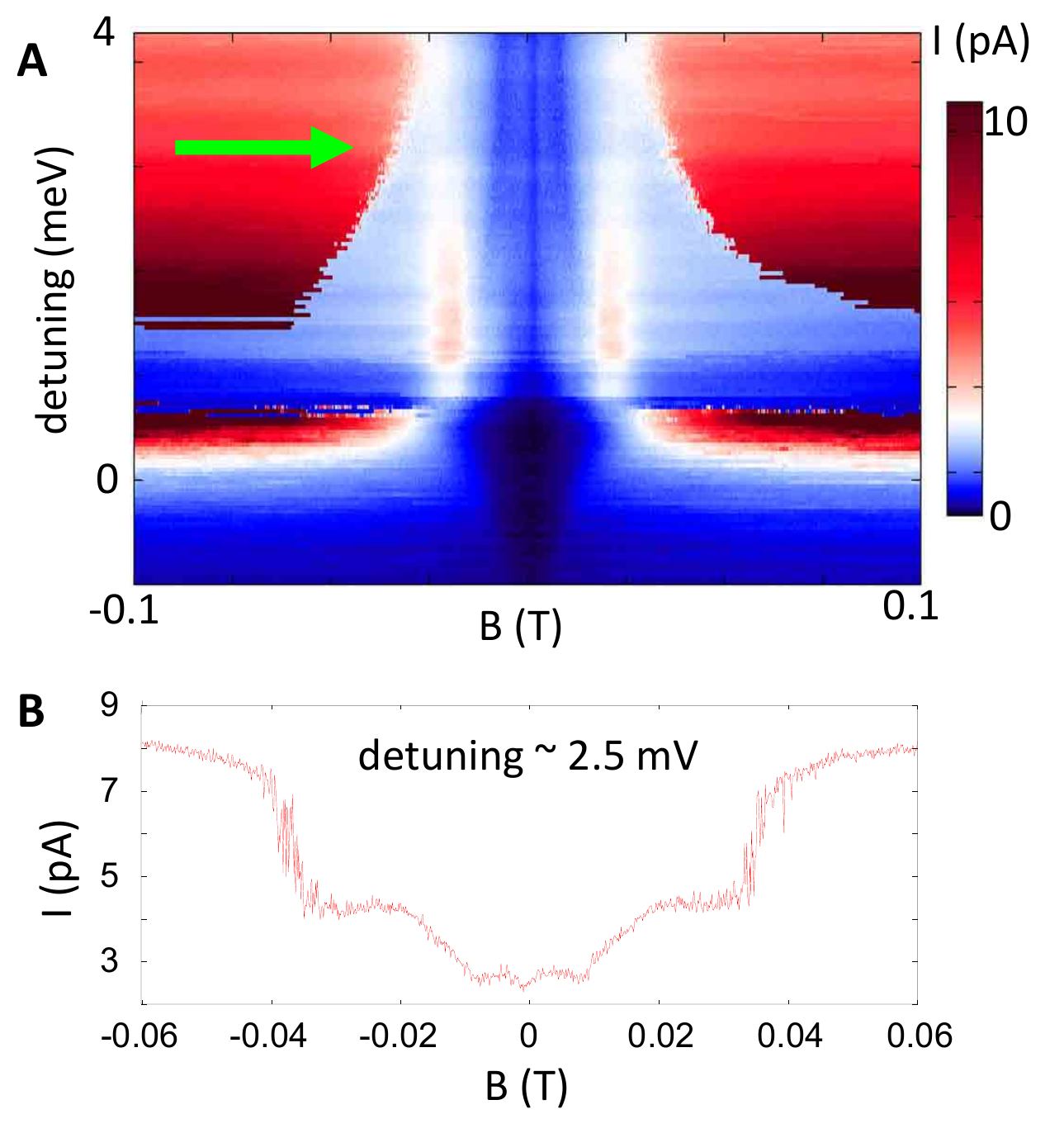}
\caption{(a) High resolution scan of the DNP onset area [device 1, (1,3)$\to$(2,2)  transition, $V_\text{dc} = -7$~mV]. In this scan, the magnetic field is swept from negative to positive and detuning is stepped after each sweep. At fields of $\approx 20$~mT and $- 20$~mT, the current exhibits local maxima with a weak detuning dependence. We interpret this as being the point where the low current state onsets. At fields lower than 20~mT the current is low due to the weakened effect of spin-orbit interaction and suppressed hyperfine mixing, see main text. In between the two 20~mT peaks, two smaller resonances appear to form a funnel shape as they move closer to zero field at higher detuning. (b) is a line cut from (a).}
\end{figure}

\begin{figure}[hp]
\includegraphics {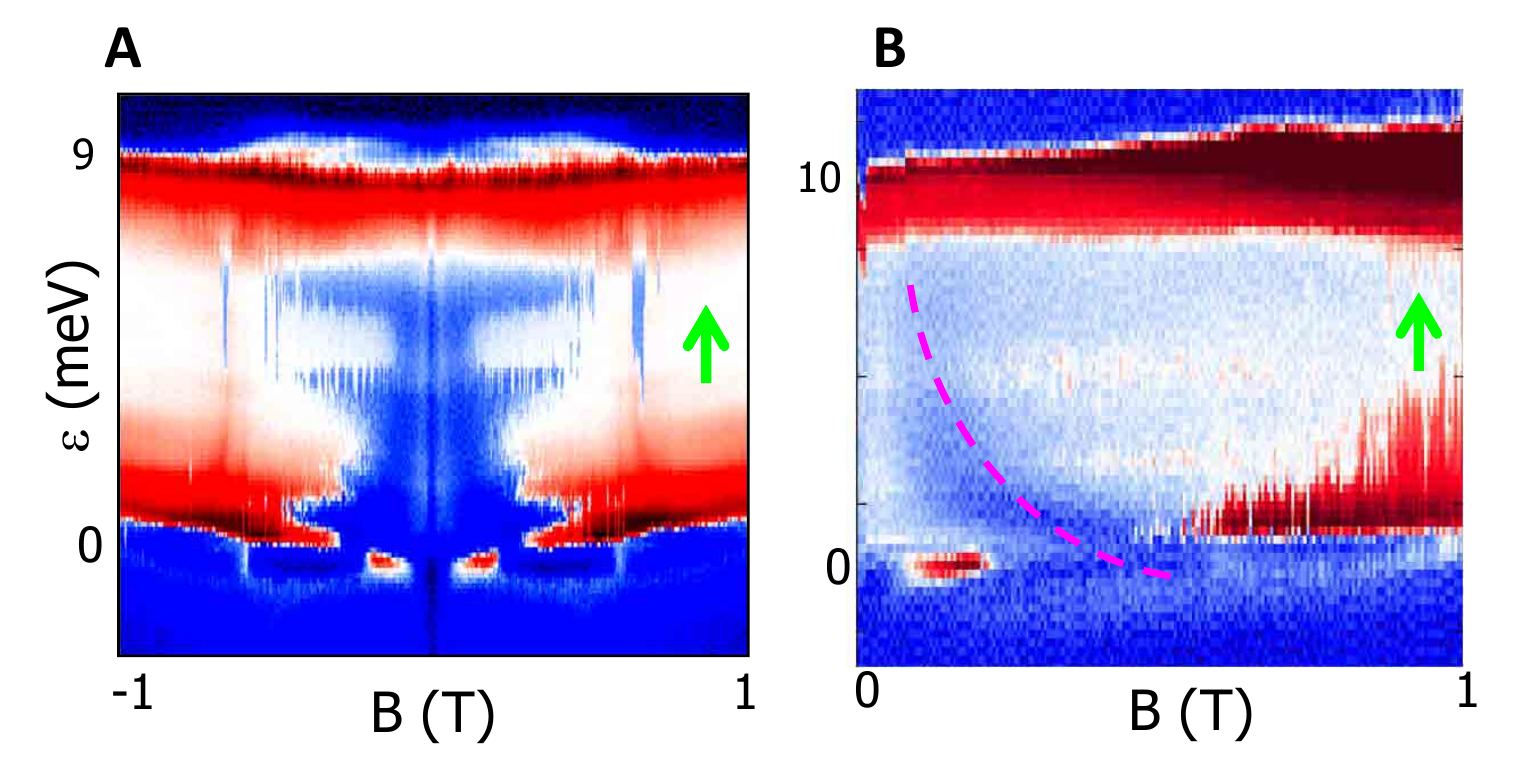}
\caption{Additional examples of DNP at the (3,1)$\to$(2,2) transition. The detuning is swept from low to high, and the field is stepped after each sweep. Panel (a) demonstrates that the maximum extent of the low current state in magnetic field can be a non-monotonic function of detuning. Panel (b) shows an even lower current state inside the low current region (dashed line). This lower current state forms a funnel shape (appears at smaller fields for higher detuning). The current is also suppressed where the funnel shape crosses the baseline near zero detuning. The origin of this state is unknown, though a $T_- \to S$ crossing is expected to have a similar shape in detuning and field.}
\end{figure}

\begin{figure}[hp]
\includegraphics {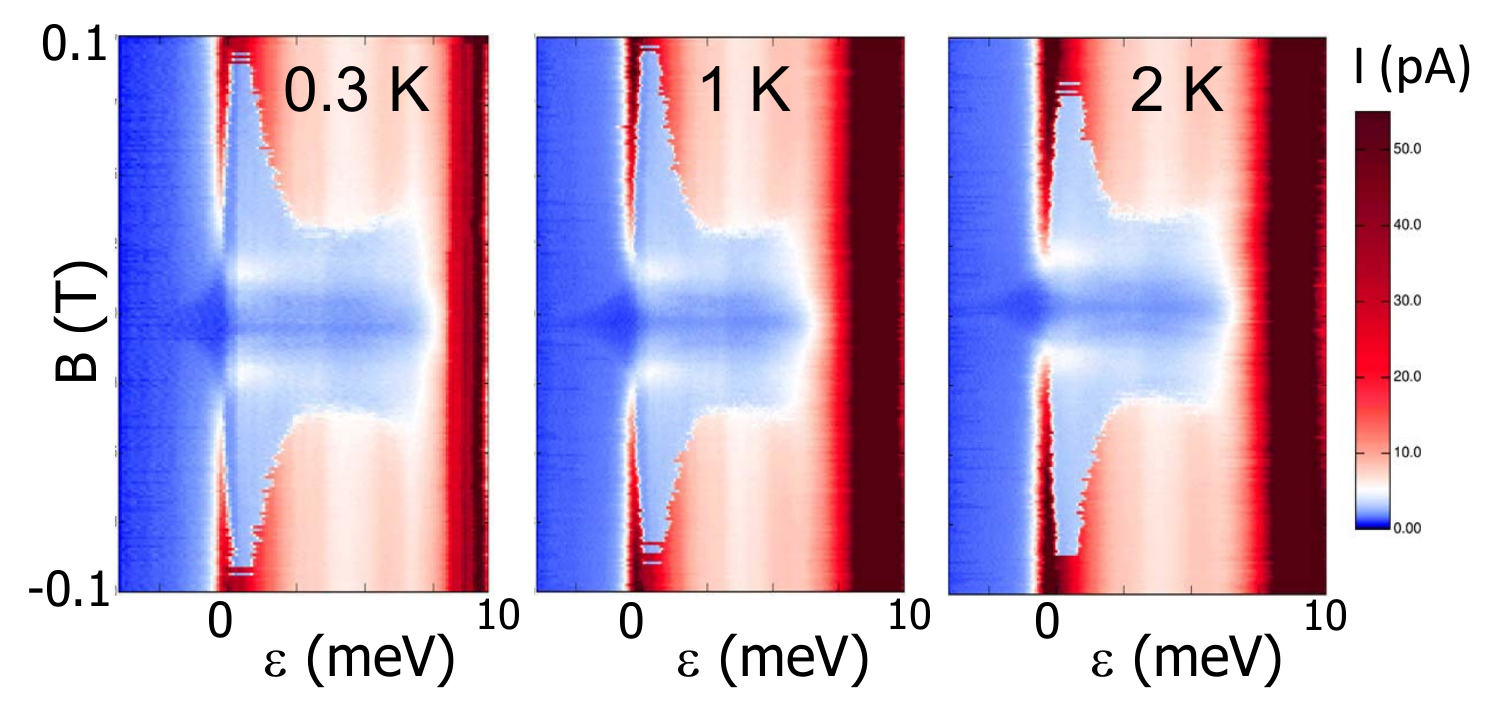}
\caption{Temperature evolution of the low current state [device 1, (1,3)$\to$(2,2) transition]. The shape of the low current state region is only slightly affected by temperature, up to 2~K. Detuning is swept from left to right.}
\end{figure}

\begin{figure}[hp]
\includegraphics[width = 15 cm]{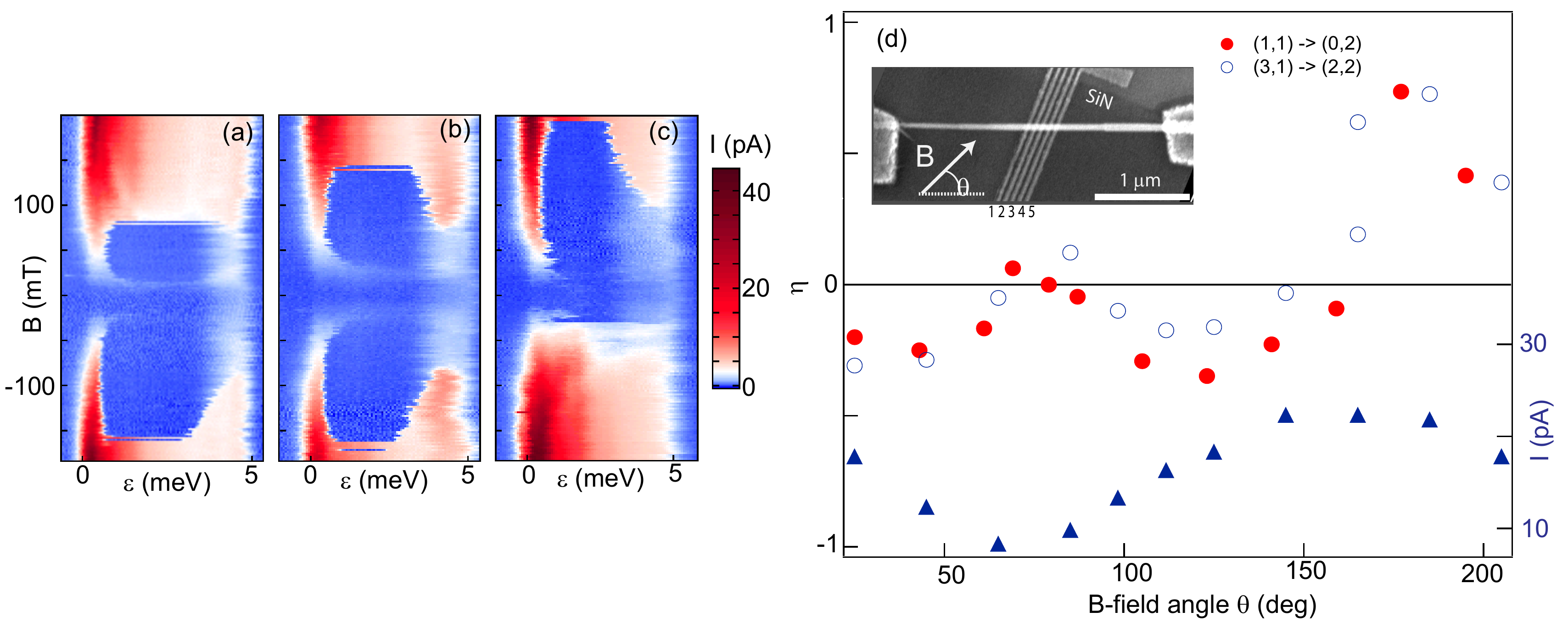}
\caption{(a) The low current state is shown to extend to different magnetic fields for positive and negative fields.  (b) When the magnetic field is oriented perpendicularly to the nanowire, the DNP pattern is nearly symmetric. (c) For fields directed along the nanowire, DNP is suppressed almost completely for one field direction. Asymmetric DNP was confirmed in device 2, where the magnetic field was fixed at an angle of 45 $\deg$ to the nanowire (Fig.\ 2). Data in Fig.\ 1 were obtained with the magnetic field perpendicular to the nanowire (out-of-plane). In that sample, pattern of the low current state is highly symmetric. In (a)--(c), the detuning is swept and the field is stepped from positive to negative for angles 25, 105 and 185 degrees with respect to nanowire. Measurements are from device 3, at the (3,1)$\rightarrow$(2,2) transition, and $V_\text{sd} = 6$~mV. The patterns are unchanged if the direction of the stepping of the magnetic field is reversed. (d) Angular dependence of the DNP asymmetry, characterized by an asymmetry parameter $\eta = (B_++ B_-)/(B_+-B_-)$, where $B_+$ and $B_-$ are the maximum positive and negative fields to which the low current state extends. The ratio $\eta$ for the (1,1)$\rightarrow$(0,2) and (3,1)$\rightarrow$(2,2) transitions (circles, left axis), and the current at $B = 200$~mT (triangles, right axis) for a range of field angles are shown. The inset shows device 3 including gates 1--5, the source (S) and drain (D) contacts, and the angle $\theta$ giving the orientation of the magnetic field. Recent theory suggests that DNP may exhibit an asymmetry due to spin-polarized injection into the double dot~\cite{RudnerRashbaPRB11}. In order to account for the apparent violation of time-reversal symmetry, the model assumes that spin-polarized injection is allowed even at zero field due to spin-orbit interaction in the nanowire leads. The model predicts maximum asymmetry when the external field is aligned with spin-injection field. The simplest interpretation of this spin-injection field is to identify it with the spin-orbit field in the nanowire. However, the direction of the spin-orbit field extracted from spin blockade anisotropy [panel (d)] in device 3 points perpendicular to the nanowire, i.e., orthogonal to the DNP asymmetry.}
\end{figure}

\section{Theoretical considerations}

\noindent As argued in the main text, the data presented in Figs.\ 1 and 2 of the main text suggest an explanation of the high and low current states in terms of small changes in the gradient $\Delta B_N^z$ of the effective nuclear fields in the two dots. The scenario proposed was that in the low current state the nuclear field gradient compensates the difference in Zeeman splittings due to the $g$-factor mismatch of the two dots. If this compensation is exact and the effective Zeeman splitting for the electrons in the dots is equal, $\Delta E_Z = 0$, then there is one fully blocked unpolarized spin-orbital (1,1) state and current is low. It was suggested that a DNP feedback mechanism stabilizes this state, explaining the large range of magnetic fields in which low current could be observed. A finite effective gradient mixes the two unpolarized (1,1) states and thus unblocks the system. A high current state could therefore correspond to a situation where the nuclear fields are unpolarized and thus $\Delta E_Z \sim |(\Delta g)\mu_BB_\text{ext}|$.

In this Section of the supplementary material, we present the details of our derivation of Eq.\ (1) in the main text and we show how we produced Fig.\ 4. The derivation involves a few simple steps, following the straightforward approach of previous related works \cite{RudnerPRL07,Danon2009,Vink2009}. We treat the hyperfine interaction between the electron and nuclear spins as a small perturbation, and we thus first focus on the electron dynamics in the spin blockade setup while ignoring the hyperfine interaction with the nuclear spins. We derive a description of the effective electron dynamics in the (1,1)-subspace, and from this we extract the quasi-stationary occupation probabilities and effective decay rates of the four (1,1) states, as well as the current through the double dot. Hyperfine induced transitions between the four (1,1) states involving electron-nuclear spin flip-flops are then evaluated with Fermi's golden rule. All allowed hyperfine transition rates are combined into an equation of motion for the nuclear polarization in the two dots. Finally, we investigate two experimentally relevant quantities: (i) the typical fluctuations of the nuclear fields around the polarized stable points are estimated from the equilibrium solution of a Fokker-Planck equation for the probability distribution of the nuclear field configurations, and (ii) we investigate the predictions of our model for the maximally achievable polarization gradient.

\subsection{Electron dynamics}

\noindent We begin by writing the Hamiltonian for the electrons involved in transport. We start with the 5-dimensional basis $\{ \ket{T_+},\ket{S},\ket{T_0},\ket{T_-},\ket{S_{02}} \}$, where the first four states are spin-orbital $(1,1)$ charge states and the fifth state $\ket{S_{02}}$ is the $(0,2)$ spin singlet state.
An externally applied magnetic field splits the energies of the $(1,1)$ states. The two spin-orbital triplets $\ket{T_+}$ and $\ket{T_-}$ will acquire a (renormalized) Zeeman energy of $\pm E_Z$. The ``unpolarized'' spin-orbital states $\ket{S}$ and $\ket{T_0}$ are coupled by the Zeeman gradient $\Delta E_Z$ over the dots. The total Zeeman term in the Hamiltonian thus reads
\begin{equation}
\hat H_Z = E_Z \big\{ \ket{T_+}\bra{T_+} - \ket{T_-}\bra{T_-} \big\} + \Delta E_Z \big\{ \ket{S}\bra{T_0} + \ket{T_0}\bra{S} \big\}.
\end{equation}
The coupling of the $(1,1)$ and $(0,2)$ spin singlets reads in the spin-orbital basis
\begin{equation}
\hat H_t = -it^- \ket{T_+}\bra{S_{02}} + t_s \ket{S}\bra{S_{02}} + it_z\ket{T_0}\bra{S_{02}} +i t^+ \ket{T_-}\bra{S_{02}} +\text{H.c.},
\end{equation}
where $t^\pm = \frac{1}{\sqrt{2}} (t_x \pm i t_y)$, with $t_x$ and $t_y$ being the coupling coefficients of $\ket{S_{02}}$ to $\ket{T_x}$ and $\ket{T_y}$, the unpolarized spin-orbital triplet states along the $x$- and $y$-axis respectively \cite{danonprb09}. The ``spin-nonconserving'' tunneling elements are typically of the same magnitude $t_{x,y,z} \sim t_\text{so} \sim \alpha t_s$, where $\alpha$ can be of the order 1 for materials with strong spin-orbit coupling \cite{danonprb09}. Finally, we describe the detuning between $\ket{S}$ and $\ket{S_{02}}$,
\begin{equation}
\hat H_\varepsilon = -\varepsilon \ket{S_{02}}\bra{S_{02}},
\end{equation}
where $\varepsilon$ is taken positive. In one view, the total Hamiltonian thus looks like
\begin{equation}
\hat H = \hat H_Z + \hat H_t + \hat H_\varepsilon = \left( \begin{array}{ccccc}
E_Z & 0 & 0 & 0 & -i t^- \\
0 & 0 & \Delta E_Z & 0 & t_s \\
0 & \Delta E_Z & 0 & 0 & it_z \\
0 & 0 & 0 & -E_Z & it^+ \\
it^+ & t_s & -it_z & -i t^- & -\varepsilon \\
\end{array}\right).
\end{equation}

To this Hamiltonian we apply a basis transformation of the two unpolarized spin-orbital states $\{\ket{S},\ket{T_0}\} \to \{\ket{c},\ket{b}\}$ as
\begin{equation}
\ket{b} = \frac{it_z\ket{S} + t_s\ket{T_0}}{\sqrt{t_s^2+t_z^2}}
\quad\text{and}\quad
\ket{c} = \frac{t_s\ket{S} + it_z\ket{T_0}}{\sqrt{t_s^2+t_z^2}}.
\end{equation}
The new basis state $\ket{c}$ is coupled to the singlet $\ket{S_{02}}$, whereas $\ket{b}$ is not. In the new basis $\{\ket{T_+},\ket{c},\ket{b},\ket{T_-},\ket{S_{02}}\}$ the Hamiltonian reads
\begin{equation}
\hat H = \left( \begin{array}{ccccc}
E_Z & 0 & 0 & 0 & -i t^- \\
0 & 0 & \Delta E_Z & 0 & \sqrt{t_s^2+t_z^2} \\
0 & \Delta E_Z & 0 & 0 & 0 \\
0 & 0 & 0 & -E_Z & it^+ \\
it^+ & \sqrt{t_s^2+t_z^2} & 0 & -i t^- & -\varepsilon \\
\end{array}\right).
\end{equation}

We now assume that the decay rate of $\ket{S_{02}}$ to the outgoing lead $\Gamma_\text{out}$ is the largest energy scale involved. This implies that we can separate time scales in the electronic dynamics, and eliminate $\ket{S_{02}}$ from the Hamiltonian. We write a $4\times 4$ Hamiltonian for the $(1,1)$ states, where the effect of the coupling to the decaying $\ket{S_{02}}$ is twofold: (i) it gives rise to exchange terms in the Hamiltonian $(\hat H_t')_{ij} = 4 T_{i2} T_{2j} \varepsilon /(\Gamma_\text{out}^2+4\varepsilon^2)$ (note that we have set $\hbar = 1$ for convenience), with $T_{a2} \equiv \braket{a|\hat H_t |S_{02}}$, and (ii) leads to decay of all coupled states with the decay rates $\Gamma_i = 4|T_{i2}|^2\Gamma_\text{out}/(\Gamma_\text{out}^2+4\varepsilon^2)$, see also \cite{danonprb09}. The effective $4\times 4$ Hamiltonian reads
\begin{equation}
\hat H_{(1,1)} =
\left( \begin{array}{cccc}
E_Z & 0 & 0 & 0 \\
0 & E_t & \Delta E_Z & 0 \\
0 & \Delta E_Z & 0 & 0 \\
0 & 0 & 0 & -E_Z \\
\end{array}\right),
\label{eq:hf2}
\end{equation}
with
\begin{equation}
E_t = \frac{4\varepsilon(t_s^2+t_z^2)}{\Gamma_\text{out}^2+4\varepsilon^2},
\end{equation}
where we assumed that $E_Z \gg 4\alpha t^2\varepsilon/(\Gamma_\text{out}^2+4\varepsilon^2)$. Indeed, in the typical regime where bistabilities are observed, the energy scale characterizing the strength of spin-orbit effects, $\sim t_\text{so}^2/\Gamma_\text{out}$, is comparable or slightly larger than the equilibrium r.m.s.\ of the nuclear fields $B_N^\text{r.m.s.} \sim 1$~mT. The Zeeman splitting $E_Z$ is typically much larger: tens of mT to several T. The four basis states of (\ref{eq:hf2}) have the effective decay rates
\begin{equation}
\Gamma_\pm \equiv \Gamma_t = \frac{1}{2}(\Gamma_x + \Gamma_y), \quad
\Gamma_c = \Gamma_s+\Gamma_z, \quad\text{and}\quad
\Gamma_b = 0,
\label{eq:decrate}
\end{equation}
with
\begin{equation}
\Gamma_a \equiv \frac{4\Gamma_\text{out} t_a^2}{\Gamma_\text{out}^2+4\varepsilon^2}.
\end{equation}

With $\hat H_{(1,1)}$ and the decay rates given as in (\ref{eq:decrate}), we can find the stationary occupation probabilities of the four states as a function of the parameters in the Hamiltonian. Diagonalization of $\hat H_{(1,1)}$ yields the set of four eigenstates $\{\ket{T_+},\ket{1},\ket{2},\ket{T_-}\}$, where the two unpolarized states read
\begin{align}
\ket{1} & = \cos \frac{\theta}{2} \ket{c} + \sin\frac{\theta}{2}\ket{b}\\
 &  =  \frac{1}{\sqrt{2}}\left(\cos\frac{\theta}{2}+\sin\frac{\theta}{2}\right)\frac{t_s+it_z}{\sqrt{t_s^2+t_z^2}}\ket{\ua\da}+ \frac{1}{\sqrt{2}}\left(\cos\frac{\theta}{2}-\sin\frac{\theta}{2}\right)\frac{-t_s+it_z}{\sqrt{t_s^2+t_z^2}}\ket{\da\ua},\nonumber \\
\ket{2} & = \cos\frac{\theta}{2} \ket{b} - \sin\frac{\theta}{2}\ket{c}\\
& = \frac{1}{\sqrt{2}}\left(\cos\frac{\theta}{2}-\sin\frac{\theta}{2}\right)\frac{t_s+it_z}{\sqrt{t_s^2+t_z^2}}\ket{\ua\da}+ \frac{1}{\sqrt{2}}\left(\cos\frac{\theta}{2}+\sin\frac{\theta}{2}\right)\frac{t_s-it_z}{\sqrt{t_s^2+t_z^2}}\ket{\da\ua},\nonumber
\end{align}
where $\theta$ is defined by $\tan\theta = 2(\Delta E_Z)/E_t$. The eigenenergies of these states are
\begin{equation}
E_1 = \frac{1}{2}E_t + \frac{1}{2}\sqrt{E_t^2 + 4(\Delta E_Z)^2}
\quad\text{and}\quad
E_2 = \frac{1}{2}E_t - \frac{1}{2}\sqrt{E_t^2 + 4(\Delta E_Z)^2},
\label{eq:eigen}
\end{equation}
and their effective decay rates read
\begin{equation}
\Gamma_1 = \left( \cos\frac{\theta}{2}\right)^2 \Gamma_c
\quad\text{and}\quad
\Gamma_2 = \left( \sin \frac{\theta}{2} \right)^2 \Gamma_c.
\label{eq:dec}
\end{equation}

We then construct a set of master equations to describe the occupation probabilities $p_a$ of the four (1,1) states. We assume that the charge transitions (0,2)$\rightarrow$(0,1)$\rightarrow$(1,1) take place on a time scale much faster than all decay rates of the (1,1) states, and that refilling of all (1,1) is equally likely. We then find the occupation probabilities
\begin{align}
p_\pm & = \frac{\Gamma_t^{-1}}{2\Gamma_t^{-1}+[(\sin \tfrac{1}{2}\theta)^{-2}+(\cos \tfrac{1}{2}\theta)^{-2}]\Gamma_c^{-1}}, \label{eq:ppm}\\
p_1 & =  \frac{(\cos \tfrac{1}{2}\theta)^{-2}\Gamma_c^{-1}}{2\Gamma_t^{-1}+[(\sin \tfrac{1}{2}\theta)^{-2}+(\cos \tfrac{1}{2}\theta)^{-2}]\Gamma_c^{-1}},\\
p_2 & = \frac{(\sin \tfrac{1}{2}\theta)^{-2}\Gamma_c^{-1}}{2\Gamma_t^{-1}+[(\sin \tfrac{1}{2}\theta)^{-2}+(\cos \tfrac{1}{2}\theta)^{-2}]\Gamma_c^{-1}}.\label{eq:p2}
\end{align}
These occupation probabilities and the eigenenergies found above are the ingredients used to plot the spectrum in Fig.\ 4(a) in the main text (thick blue lines). Under the assumptions made above, the current through the double dot simply follows as
\begin{equation}
I = \frac{4e}{2\Gamma_t^{-1}+[(\sin \tfrac{1}{2}\theta)^{-2}+(\cos \tfrac{1}{2}\theta)^{-2}]\Gamma_c^{-1}},
\label{eq:curr}
\end{equation}
on average 4 electrons are transported in a time which is the sum of the four decay times of the (1,1) states. This expression was used to produce the dashed brown plot in Fig.\ 4(a).

\subsection{Dynamic nuclear polarization}

\noindent We now include hyperfine interaction into the picture. The nuclear spins in both dots, represented by the operators $\mathbf{\hat I}_{L,R}$, are coupled to the electron spins via hyperfine interaction,
\begin{equation}
\hat H_{\text{hf}} = \frac{A}{2N_L} \sum_{k_L} \left\{ 2\hat S_L^z \hat I^z_{L,k} + \hat S^+_L \hat I^-_{L,k} + \hat S^-_L\hat I^+_{L,k} \right\}
+ \frac{A}{2N_R} \sum_{k_R} \left\{ 2\hat S_R^z \hat I^z_{R,k} + \hat S^+_R \hat I^-_{R,k} + \hat S^-_R\hat I^+_{R,k} \right\},
\label{eq:hhf}
\end{equation}
where the two sums run over all nuclei in the two dots. For simplicity we assume that all nuclear spins are equally strongly coupled to the electron spin, which reduces the prefactor to the hyperfine coupling energy $A$ divided by an effective number of nuclei $N_{L,R}$ in the dots.

The $z$-components of the coupling in (\ref{eq:hhf}) lead to an Overhauser shift of the Zeeman energy of the electrons in the two dots. Including this Overhauser shift into our description is done by setting $E_Z = \bar g \mu_B B_\text{ext} + \tfrac{1}{2}A[\langle \hat I^z_L \rangle + \langle \hat I^z_R \rangle]$ and $\Delta E_Z = (\Delta g)\mu_B B_\text{ext} + \tfrac{1}{2}A[\langle \hat I^z_L \rangle - \langle \hat I^z_R \rangle]$. In terms of the degree of nuclear polarization of the two dots, $-1< P_{L,R}<1$, we can write $E_Z = \bar g \mu_B B_\text{ext} + AIP$ and $\Delta E_Z = (\Delta g)\mu_B B_\text{ext} + AI(\Delta P)$ introducing the average dot polarization $P = \tfrac{1}{2}(P_L+P_R)$ and polarization gradient $\Delta P = \tfrac{1}{2}(P_L-P_R)$.

The hyperfine flip-flop terms $\hat S^+\hat I^-$ and $\hat S^-\hat I^+$ are the ones that can cause spin exchange between the electrons and the nuclei and could lead to DNP. In the main text we explained in qualitative terms how the dominant hyperfine flip-flop processes close to the point $\Delta E_Z = 0$ lead to stabilization of this point. We now have all ingredients at hand to evaluate the DNP rates explicitly. We employ Fermi's golden rule to calculate all flip-flop rates, and we add for each dot all rates flipping nuclear spin up and all rates flipping nuclear spin down,
\begin{align}
R^+_d = \frac{A^2}{4N_d}\frac{1}{E_Z^2}p_{\da,d}\Big\{ & \Gamma_1p_+|{\bra{1}\hat S^-_d\ket{T_+}}|^2 + \Gamma_2p_+|{\bra{2}\hat S^-_d\ket{T_+}}|^2 \nonumber\\
& + \Gamma_tp_1|{\bra{T_-}\hat S^-_d\ket{1}}|^2 + \Gamma_tp_2|{\bra{T_-}\hat S^-_d\ket{2}}|^2 \Big\},\label{eq:rpd}\\
R_d^- = \frac{A^2}{4N_d}\frac{1}{E_Z^2}p_{\ua,d}\Big\{ & \Gamma_1p_-|{\bra{1}\hat S^+_d\ket{T_-}}|^2 + \Gamma_2p_-|{\bra{2}\hat S^+_d\ket{T_-}}|^2 \nonumber\\
& + \Gamma_tp_1|{\bra{T_+}\hat S^+_d\ket{1}}|^2 + \Gamma_tp_2|{\bra{T_+}\hat S^+_d\ket{2}}|^2 \Big\},\label{eq:rmd}
\end{align}
where $p_{\ua(\da),d}$ is the fraction of nuclear spins in dot $d$ which has its spin up(down) \footnote{We assume for simplicity nuclear spin 1/2. Changing to the correct nuclear spin involves only the inclusion of a numerical prefactor to the pumping equations.}, and we used that $E_Z \gg \Gamma_{1,2,t}$. The occupation probabilities $p_{1,2,\pm}$ used in (\ref{eq:rpd}) and (\ref{eq:rmd}) are the ones found in the previous Subsection, (\ref{eq:ppm})--(\ref{eq:p2}). These probabilities do not include the effect of the hyperfine decay rates themselves, but since we have in the regime of interest at most one blocked state, all leading order hyperfine induced corrections the the probabilities are of the order $R^\pm_d/\min\{\Gamma_t,\Gamma_c\}$, which we assume to be small.

We then combine the rates into equations of motion for the polarization in the two dots,
\begin{align}
\frac{dP_L}{dt} & = \frac{2}{N_L}(R^+_L- R^-_L) = \frac{A^2}{16 N_L^2 E_Z^2}\frac{\Gamma_c^2\sin^2 \theta +4 \Gamma_t^2}{\Gamma_c\sin^2 \theta +2 \Gamma_t} (-2P_L - \sin 2\theta),\label{eq:gl}\\
\frac{dP_R}{dt} & = \frac{2}{N_R}(R^+_R- R^-_R) = \frac{A^2}{16 N_R^2 E_Z^2}\frac{\Gamma_c^2\sin^2 \theta +4 \Gamma_t^2}{\Gamma_c\sin^2 \theta +2 \Gamma_t} (-2P_R + \sin 2\theta),\label{eq:gr}
\end{align}
where we used that $p_{\ua(\da),d} = \tfrac{1}{2}(1\pm P_d)$. We finally rearrange (\ref{eq:gl}) and (\ref{eq:gr}) into time-evolution equations for the average polarization and polarization gradient,
\begin{align}
\frac{d(\Delta P)}{dt} & = \frac{A^2}{32 E_Z^2} \frac{\Gamma_c^2\sin^2 \theta +4 \Gamma_t^2}{\Gamma_c\sin^2 \theta +2 \Gamma_t} \left\{ -\frac{N_L^2+N_R^2}{N_L^2N_R^2}[2(\Delta P) + \sin 2\theta]+ \frac{N_L^2-N_R^2}{N_L^2N_R^2}2P\right\},\label{eq:pumpdp}\\
\frac{dP}{dt} & = \frac{A^2}{32 E_Z^2} \frac{\Gamma_c^2\sin^2 \theta +4 \Gamma_t^2}{\Gamma_c\sin^2 \theta +2 \Gamma_t} \left\{ -\frac{N_L^2+N_R^2}{N_L^2N_R^2}2P + \frac{N_L^2-N_R^2}{N_L^2N_R^2}[2(\Delta P) + \sin 2\theta]\right\}.\label{eq:pumptp}
\end{align}
In the limit of equal numbers of spinful nuclei in the two dots, $N_L = N_R \equiv N$, this set of equations reduces to
\begin{align}
\frac{d(\Delta P)}{dt} & = -\frac{A^2}{16 N^2 E_Z^2} \frac{\Gamma_c^2\sin^2 \theta +4 \Gamma_t^2}{\Gamma_c\sin^2 \theta +2 \Gamma_t} [2(\Delta P) + \sin 2\theta]-\frac{1}{\tau}(\Delta P),\label{eq:pumpdp2}\\
\frac{dP}{dt} & = -\frac{A^2}{8 N^2 E_Z^2} \frac{\Gamma_c^2\sin^2 \theta +4 \Gamma_t^2}{\Gamma_c\sin^2 \theta +2 \Gamma_t}P-\frac{1}{\tau}P,\label{eq:pumptp2}
\end{align}
where we included a phenomenological relaxation rate $1/\tau$, with $\tau$ being typically very long (often on the scale of seconds). A deviation from $N_L = N_R$ leads to small corrections to the pumping curves and stable points. To leading order, one can expect corrections of relative magnitude $(N_L^2-N_R^2)^2/(N_L^2+N_R^2)^2$, which is typically very small.

Eq.\ (2) in the main text is (\ref{eq:pumpdp2}), where for simplicity we have set $t_x = t_y = t_z = t_\text{so}$. Within this approximation, we have for the function $f(\theta)$ introduced in the main text
\begin{equation}
f(\theta) = \frac{(t_s^2+t_\text{so}^2)^2\sin^2 \theta +4 t_\text{so}^4}{t_s^2(t_s^2+t_\text{so}^2)\sin^2 \theta +2 t_s^2t_\text{so}^2}
=\frac{(1+\alpha^2)^2\sin^2 \theta +4 \alpha^4}{(1+\alpha^2)\sin^2 \theta +2 \alpha^2}.
\end{equation}
Eq.\ (\ref{eq:pumpdp2}) is the one we used to produce the plots in Fig.\ 4(b) of the main text. As already can be seen from the plots, for a large range of magnetic fields, the DNP mechanism can indeed stabilize the system close to the point with $\Delta E_Z = 0$. 

Let us briefly comment here on the differences between the mechanism proposed in this work and previously investigated mechanisms. The key difference compared to similar systems hosted in GaAs, is the strong spin-orbit coupling, effectively lifting the blockade of the Zeeman split-off triplet states. As long as $E_Z > \max\{(t_\text{so}^2/\Gamma_\text{out}), \Delta E_Z \}$ we have at most one blocked state [see Fig.\ 4(a) of the main text]. This allows us to neglect the hyperfine assisted escape rates when calculating the $p_i$ [the thickness of the lines in Fig.\ 4(a) of the main text]. In a similar situation in GaAs, the hyperfine rates themselves could be the dominant escape rates out of the three triplet states, thereby heavily influencing the $p_i$, which could in turn significantly affect the DNP \cite{RudnerPRB11}. Other groups used elaborate gate pulsing schemes to drive one specific hyperfine transition, usually isolating the $S\to T_+$ transition \cite{ReillyScience08,FolettiNatPhys09}. In this case, a small imbalance of the $\ua\da$- and $\da\ua$-components in the electronic ground state can also cause a significant pumping of the nuclear field gradient, ultimately leading to a narrowing of the distribution function of the nuclear field gradient around $\Delta P=0$ \cite{PhysRevB.81.041304}. This narrowing is accompanied by an overall drift of both nuclear polarizations to lower values since all nuclear spin flips have the same preferred direction. Apart from this, when driving the $S\to T_+$ transition, a small asymmetry in the number of nuclei in the two dots ($N_L \neq N_R$) can also have a dramatic effect: the resulting imbalance in the spin-flip rates could lead to continuous increasing of the field gradient. In our setup, a small asymmetry between $N_L$ and $N_R$ only causes small corrections to the pumping equations (\ref{eq:pumpdp}) and (\ref{eq:pumptp}).

\subsection{Maximum gradient achievable}

\noindent A useful quantity to extract from Eq.\ (\ref{eq:pumpdp2}) is the maximally achievable polarization gradient $(\Delta P)_\text{max}$. Using the fact that the gradient at the stable point exactly cancels the field gradient caused by the $g$-factor mismatch, we can relate this maximum as $AI(\Delta P)_\text{max} = -(\Delta g)\mu_B B_\text{max}$, where $B_\text{max}$ is the maximum field for which a bistability could be observed. We can find this maximum from zooming in on the peak-dip structure. We see that the top \footnote{We focus here for definiteness on {\it negative} $E_Z$, which corresponds to a {\it positive} applied magnetic field, leading to a {\it positive} gradient $\Delta P$ at the stable point, assuming that $g_L>g_R$.} of the DNP peak corresponds with $\sin 2\theta \approx -1$, so that $\theta \approx -\pi/4$. This makes $\sin^2 \theta \approx 1/2$, so pumping at this local maximum reads
\begin{equation}
\left.\frac{d(\Delta P)}{dt}\right|_\text{max} = -\frac{A^2}{4 N^2E_Z^2}\frac{\Gamma t_s^2}{\Gamma^2+4\varepsilon^2}\frac{N^2\Gamma^2}{r t_s^2\tau}\left\{ -1 + 2(\Delta P)\right\} -\frac{1}{\tau}(\Delta P),
\label{eq:dpmax}
\end{equation}
where we introduced the coefficient
\begin{equation}
r = f\left(-\frac{\pi}{4}\right)\frac{t_s^2\tau}{N^2\Gamma} \approx \frac{(3\alpha^2+1)^2}{5\alpha^2+1}\frac{t_s^2\tau}{N^2\Gamma},
\end{equation}
for convenience of notation. When this local maximum in the pumping curve is exactly zero, we know that the corresponding polarization gradient $(\Delta P)_\text{max}$ is the maximum $\Delta P$ achievable, and from this we find $B_\text{max}$, the largest field for which in principle bistabilities could be observed. Setting $(\Delta P)_\text{max} = 0$ yields 
\begin{equation}
|\bar g \mu_B B_\text{max}| = \frac{|\bar g|}{|\Delta g|} \frac{IA^3}{2A^2+4r[1+4(\varepsilon/\Gamma_\text{out})^2](\bar g \mu_B B_\text{max})^2},
\label{eq:bmax}
\end{equation}
which is a cubic equation and can easily be solved.

\begin{figure}[t]
\begin{center}
\includegraphics[width=85mm]{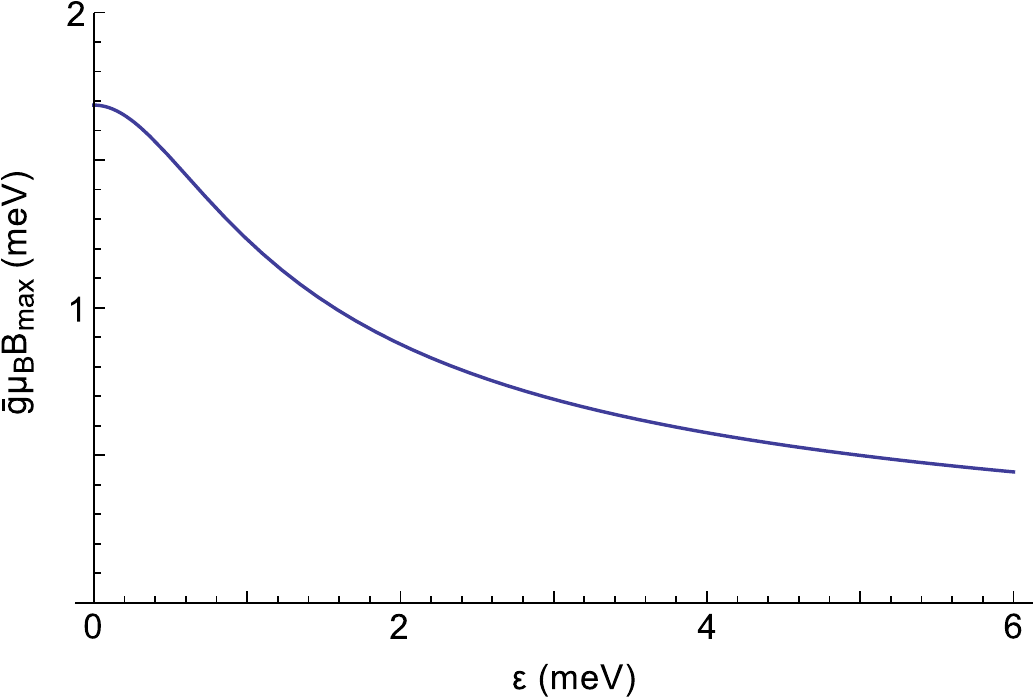}
\caption{The estimated maximum field for which the pumping curve (\ref{eq:pumpdp2}) yields a stable point at $\Delta E_Z = 0$, as a function of the detuning $\varepsilon$. We used the same parameters as in Fig.\ 4b in the main text: $AI = 0.7$ meV, $I = 1/2$, $N = 10^5$, $\Gamma_\text{out} = 1.5$ meV, $t_s = 60~\mu$eV, $t_\text{so} = 40~\mu$eV, $(\Delta g)/\bar g = 0.05$, and $1/\tau = 3\cdot 10^{-10}~\mu$eV.}\label{fig:supfig}
\end{center}
\end{figure}

In Fig.\ \ref{fig:supfig} we plot the dependence of $B_\text{max}$ on the detuning $\varepsilon$, using the same parameters as were used in Fig.\ 4(b) in the main text. We see that for large detuning the range of magnetic field in which hysteresis can be observed is suppressed. From Eq.\ (\ref{eq:bmax}) we find that the large-$\varepsilon$ behavior of the maximal field is $B_\text{max} \propto \varepsilon^{-2/3}$. Boundaries of the hysteretic regime in the $(B,\varepsilon)$-plane with similar shape have indeed been observed in experiment, see Figs.\ S1--S6 but also for instance the data presented in \cite{pfundprl07}.

\subsection{Fluctuations of the nuclear fields around the stable points}

\noindent To estimate the typical fluctuations of $\Delta P$ around the stable point with $\Delta E_Z = 0$, we largely follow the approach of \cite{PhysRevLett.100.056603,Vink2009}. Since the dynamics found in (\ref{eq:pumpdp2}) and (\ref{eq:pumptp2}) for $\Delta P$ and $P$ are independent, we can focus on the fluctuations of $\Delta P$ separately. We label all possible configurations of the nuclear spins in the two dots resulting in different $\Delta P$ by $n = \tfrac{1}{2}(N_{+,L} - N_{-,L}) - \tfrac{1}{2}(N_{+,R}-N_{-,R})$, where $N_{\pm,L(R)}$ denotes the number of nuclear spins in the left(right) dot with spin up or down. This labeling results in $2N$ different discrete values for $n$, ranging from $-N$ to $N$. A nuclear spin flip in either of the dots results in a change of $n$ by 1. We now construct a Fokker-Planck equation for the probability distribution $\mathcal{P}(n)$, based on the master equation
\begin{equation}
\frac{\partial \mathcal{P}(n)}{\partial t} = -\mathcal{P}(n) [ \Gamma_+(n) + \Gamma_-(n) ] + \mathcal{P}(n-1) \Gamma_+(n-1) + \mathcal{P}(n+1) \Gamma_-(n+1).
\label{eq:master}
\end{equation}
Here, $\mathcal{P}(n)$ gives probability to find the system in the configuration $n$, and $\Gamma_\pm(n)$ is the rate at which the nuclear spin baths flip from the configuration $n$ to $n\pm 1$. Using the fact that the number of configurations $2N$ is large, we can go over to the continuous limit, yielding
\begin{equation}
\frac{\partial \mathcal{P}}{\partial t} = \frac{\partial}{\partial n} \left\{(\Gamma_--\Gamma_+)\mathcal{P} +\frac{1}{2} \frac{\partial}{\partial n} (\Gamma_-+\Gamma_+)\mathcal{P} \right\},
\label{eq:fp}
\end{equation}
which indeed is a Fokker-Planck equation.

Since the functions $\Gamma_\pm (n)$ are smooth on the scale of $n$, we can use that $|\partial_n \Gamma_\pm (n) | \ll \Gamma_\pm$. In terms of the polarization gradient $\Delta P \equiv n/N$ this allows us to find the approximate solution valid close to the stable point $\Delta P = \Delta P_0$,
\begin{equation}
\mathcal{P}(\Delta P) \approx \mathcal{P}(\Delta P_0) \exp \Bigg\{ -\frac{(\Delta P-\Delta P_0)^2}{2\sigma^2} \Bigg\},
\label{eq:appr}
\end{equation}
where the coefficient $\sigma^2$ is defined as
\begin{equation}
\sigma^2 = \frac{1}{2N} \left( -\frac{\partial}{\partial (\Delta P)} \left. \frac{\Gamma_+-\Gamma_-}{\Gamma_++\Gamma_-}\right|_{\Delta P_0} \right)^{-1} = \frac{1}{2N} \left. \frac{\Gamma_++\Gamma_-}{\frac{\partial}{\partial (\Delta P)} (\Gamma_--\Gamma_+)}\right|_{\Delta P_0}.
\label{eq:sigmas}
\end{equation}
In the last simplification, we used the fact that $\left. (\Gamma_+ - \Gamma_- )\right|_{\Delta P_0} = 0$. We see that this solution for $\mathcal{P}(\Delta P)$ is a Gaussian distribution with a width $\sigma$.

The rates of change of $n$ can be expressed in terms of the spin flip rates as $\Gamma_+ = R_L^+ + R^-_R$ and $\Gamma_- = R_L^- + R_R^+$, from which we can write an explicit expression for $\sigma$. To arrive at a rough estimate, we evaluate the fluctuations at the point where exactly $\theta = 0$. This yields
\begin{equation}
\sigma^2 \approx \frac{1}{2N}
\frac{E_t[N^2E_Z^2(\Gamma_\text{out}^2+4\varepsilon^2)+ \alpha^2 A^2\Gamma_\text{out}t_s^2\tau]}{E_t[N^2E_Z^2(\Gamma_\text{out}^2+4\varepsilon^2) +\alpha^2 A^2\Gamma_\text{out}t_s^2\tau] + 2\alpha^2 A^3I \Gamma_\text{out}t_s^2\tau}.
\label{eq:sigma2}
\end{equation}
In reality, $\theta$ is not exactly zero at the stable point. Expanding (\ref{eq:sigmas}) around the point where $\theta = 0$, we find to leading order corrections of $\sim \delta(\Delta P)[(\Delta g)\mu_BB_\text{ext}/E_t]$ to both $(\Gamma_+ + \Gamma_-)$ and $\partial_{(\Delta P)}(\Gamma_- - \Gamma_+)$, where $\delta(\Delta P)$ is the deviation of $\Delta P$ from the point with $\theta = 0$. Due to the shape of the pumping curve, $\theta$ will always lie within $-\pi/4 < \theta < \pi/4$, which sets the scale for the maximal $\delta(\Delta P)$. From this estimate, it follows that both relative corrections are of the order $\lesssim |(\Delta g)\mu_B B_\text{ext}/AI|$, which is indeed small.

In the regime where
\begin{equation}
AI \gg E_t
\quad\text{and}\quad
\frac{AI}{E_t}\frac{A^2}{N^2E_Z^2}\frac{\Gamma_\text{out} t_\text{so}^2}{\Gamma_\text{out}^2+4\varepsilon^2} \gg \frac{1}{\tau},
\label{eq:cond}
\end{equation}
the fluctuations can be estimated as
\begin{equation}
\sigma^2 \approx \frac{\varepsilon(t_s^2+t_\text{so}^2)}{NAI(\Gamma_\text{out}^2+4\varepsilon^2)}\left(1+\frac{N^2E_Z^2}{\tau A^2}\frac{\Gamma_\text{out}^2+4\varepsilon^2}{\Gamma_\text{out} t_\text{so}^2}\right).
\end{equation}
For all experimentally relevant parameters we expect (\ref{eq:cond}) to hold.

If we combine these estimates for the fluctuations of the nuclear field gradient with the simple expression we found for the double dot current (\ref{eq:curr}), we can investigate what our model predicts for the current fluctuations induced by the nuclear field fluctuations. At first sight, one might expect lower current noise in a state with small fluctuations. However, since the current is a very sensitive function of $\Delta E_Z$ close to the polarized stable point, small fluctuations around $\theta = 0$ can have a dramatic effect on the current. Expanding the current around $\theta=0$ up to second order in $\delta(\Delta P)$, we can estimate the magnitude of the current fluctuations around the polarized stable point as
\begin{equation}
\delta I \approx 4e\frac{A^2I^2\Gamma_\text{out}}{E_t\varepsilon}\sigma^2,
\end{equation}
which under the condition of (\ref{eq:cond}) simplifies to
\begin{equation}
\delta I \approx e\frac{AI\Gamma_\text{out}}{N\varepsilon}\left(1+\frac{N^2E_Z^2}{\tau A^2}\frac{\Gamma_\text{out}^2+4\varepsilon^2}{\Gamma_\text{out} t_\text{so}^2}\right).
\end{equation}
To put this in perspective, we evaluate the current for $\Delta P \to \pm \infty$, i.e., for $\theta \to \pm \pi/2$,
\begin{equation}
I_\text{max} = e\frac{2\Gamma_\text{out}E_t}{\varepsilon}\frac{1+\alpha^2}{1+3\alpha^2}.
\end{equation}
This allows us to express the relative fluctuations in the current: the magnitude of the low current fluctuations scaled to the difference in high and low current, $I_\text{max}$. We thus find
\begin{equation}
\frac{\delta I}{I_\text{max}} \approx \frac{AI}{2NE_t}\frac{1+3\alpha^2}{1+\alpha^2}\left(1+\frac{N^2E_Z^2}{\tau A^2}\frac{\Gamma_\text{out}^2+4\varepsilon^2}{\Gamma_\text{out} t_\text{so}^2}\right).
\label{eq:ii}
\end{equation}

As written above, the typical regime where strong bistabilities are observed is that of strong, but not too strong, coupling, where the energy scale characterizing the strength of spin-orbit effects in the (1,1)-subspace is comparable or slightly larger than the equilibrium r.m.s.\ value of the nuclear fields. More quantitatively, we expect to be in the regime where $t_\text{so}^2/\Gamma_\text{out} \gtrsim B_N^\text{r.m.s.} \sim AI/\sqrt{N}$. We see that when $\alpha \sim 1$ (strong spin-orbit coupling), we have $E_t \sim t_\text{so}^2/\Gamma_\text{out}$ and thus find for the prefactor in (\ref{eq:ii})
\begin{equation}
\frac{AI}{2NE_t}\frac{1+3\alpha^2}{1+\alpha^2}  \lesssim \frac{1}{\sqrt{N}},
\end{equation}
which is small, typically $\sim 10^{-2}$--$10^{-3}$. The actual ratio $\delta I/I_\text{max}$ then depends on the second term within the brackets, which is $\sim 1$--10 for the parameters used in the main text. In this regime the relative current fluctuations are thus indeed small.

This ratio however, does not say anything about the relative magnitude of the fluctuations around $\Delta P=0$ and those around the polarized state. In fact, for the parameters we used in the main text, we find that the model predicts the current fluctuations in the low current state to be larger than in the high current state, which is consistent with Fig.\ 1(c) of the main text. Other data sets however (such as shown in  Fig.\ S2) seem to indicate opposite behavior. A full understanding of these different observations requires a much more quantitative examination of all parameters which goes beyond the scope of this work. We just note that the simple model presented here in fact does not exclude the possibility of having larger current fluctuations in the high current state.

\bibliographystyle{apsrev}
\bibliography{sb}{}
\end{document}